\documentstyle[12pt]{article}
 \newcommand{\bea}{\begin{equation}}
 \newcommand{\eea}{\end{equation}}
 \newcommand{\ber}{\begin{eqnarray}}
 \newcommand{\eer}{\end{eqnarray}}
\begin{document}
\title{SHEAR THINNING OF A CRITICAL VISCOELASTIC FLUID}
\author{Palash Das and Jayanta K.Bhattacharjee\\
Department of Theoretical Physics \\
Indian Association for the Cultivation of Science \\
Jadavpur, Kolkata 700 032, India}
\date{}
\maketitle
\begin{abstract}
The frequency and shear dependent critical viscosity at a
correlation length $\xi=\kappa^{-1}$, has the form
$\eta=\eta_{0}\kappa^{-x_{\eta}}G(z_{1},z_{2})$, where $z_{1}$ and
$z_{2}$ are the independent dimensionless numbers in the problem
defined as $z_{1}=\frac{-i\omega}{2\Gamma_{0}\kappa^{3}}$ and
$z_{2}=\frac{-i\omega}{2\Gamma_{0}\kappa_{c}^{3}}$. The decay rate
of critical fluctuations of correlation length $\kappa^{-1}$ is
$\Gamma_{0}\kappa^{3}$ and $k_{c}$ is the effective wave number
for which $\Gamma_{0}k_{c}^{3}=S$, the shear rate. The function
$G(z_{1},z_{2})$ is calculated in a one loop self-consistent
theory.\vspace{1cm}

PACS number(s):64.60Ht
\end{abstract}
\newpage
Response functions in general tend to diverge near a second order
phase transition point. As the fluctuations become more long
ranged and drive the system towards the transition, the response
functions in the thermodynamic limit (if static) and in the
hydrodynamic limit (if dynamic) diverge. Some like the
susceptibility (static) and thermal diffusivity (dynamic) [1]
diverge strongly with an exponent one, others like the specific
heat (static) and shear viscosity (dynamic) [2-4] diverge weakly
with an exponent close to zero. The small exponents have always
provided an interesting challenge for theorists and
experimentalists. The shear viscosity of ordinary fluids or binary
mixtures, in particular, has been a favourite candidate for
pushing theory and experiment to the limit. The exponent is small
but over the last two decades the experiments [5,6] have increased
the accuracy of the measurements so strongly that theorists have
had to worry about higher loops [7,8], something which is rarely
done.
\par
The critical divergence is masked if the system is not in the
hydrodynamic limit. Hydrodynamic limit implies the wavelength is
larger than the critical correlation length and all frequencies
are lower than the rate of decay of critical fluctuations. While
wavelengths are always larger than the critical correlation
length, critical slowing down may imply that frequencies may not
be smaller than the fluctuation relaxation rate. The frequency of
an oscillating viscometer used to measure the critical viscosity
can in that case determine the viscosity of the sample and the
fluid will become viscoelastic.
\par
The frequency dependence of the viscosity predicted [9,10] and
measured over the last two decades. It has been recognized by
Oxtoby that a shear rate will also introduce a time scale the
problem and if the fluctuations are more long lived than this time
scale, then experiments on this time scale will not feel the
longest-lived fluctuations and the viscosity will be limited by
the shear rate - a phenomenon known shear thinning. The dependence
of various static quantities on the shear rate as well as its
effect on light scattering has been considered by Onuki [12] as
well as Onuki and Kawasaki [13]. With so much information
available on both theoretical and experimental fronts, experiments
have been designed to test the shear thinning of critical
viscoelastic fluid. We have carried out a one loop calculation of
the frequency and shear dependent critical viscosity and present
our results in this communication.
\par
In the hydrodynamic regime, characterized by the sole length scale
- the correlation length
$\xi=\kappa^{-1}=\xi_{0}[\frac{T-T_{c}}{T_{c}}]^{-\nu}$, with
$\nu\simeq \frac{2}{3}$, the shear viscosity diverges as \bea
\eta(\kappa)=\eta_{0}\kappa^{-x_{\eta}} \eea where $x_{\eta}$ is
the small exponent discussed above which is found to be around
0.068. The decay time of the fluctuations, $\tau$ , is given by
\bea \tau^{-1}=\Gamma(k,\kappa)=\frac{L}{\chi}k^{2} \eea where $L$
is the Onsager coefficient and $\chi$ is the static
susceptibility. The critical viscosity can , to a very good
accuracy , be taken as $\chi^{-1}(k,\kappa)=k^{2}+\kappa^{2}$ and
the Onsager coefficient or thermal diffusivity diverges at the
critical point. The behavior of wavelength and correlation length
dependent diffusivity is governed by the Kawasaki function [1]
which was simplified for practical use by Ferrell and can be used
in the form \bea L=\Gamma_{0}(k^{2}+\kappa^{2})^{-\frac{1}{2}}
\eea in $D=3$, where D is the spatial dimensionality. The
characteristic decay time of a fluctuation of wavelength equal to
the correlation length is given by \bea
\tau^{-1}=\Gamma(\kappa)=\Gamma_{0}\kappa^{3} \eea Hydrodynamic
regime implies that frequencies are such that $\omega\tau\ll1$.
However, at a fixed frequency $\omega$, as $\kappa$ decreases on
approaching the critical point, $\tau$ diverges and it is not
possible to satisfy $\omega\tau<1$. In that situation Eq.(1) can
not hold and if $\tau^{-1}\simeq0$, the response is limited by the
frequency $\omega$. Since $\omega$ scales as $\kappa^{3}$, it is
clear that the limiting viscosity will be of the form \bea
\eta(\omega)=\eta_{0}(-i\omega)^{-\frac{x_{\eta}}{3}} \eea and the
full viscosity will be governed by \bea
\eta(\kappa,\omega)=\eta_{0}\kappa^{-x_{\eta}}F(z)^{-\frac{x_{\eta}}{3}}
\eea where $F(z)$ is a function of the dimensionless variable
$z=\frac{-i\omega}{2\Gamma_{0}\kappa^{3}}$. For $z\rightarrow0$,
$F(z)\rightarrow1$ and if $z\gg1$, $F(z)\propto z$. The simplest
possible functional representation of $F(z)$ is \bea F(z)=1+\beta
z \eea Now, $\beta$ is a number of $O(1)$, which can be found from
the small z form of the one loop integral or from the $z\gg1$
form. If determined from the low frequency end
$\beta=\frac{3\pi}{16}\simeq 0.59$. From the high frequency end
$\beta \simeq 0.2$. The two loop calculation is found to enhance
$\beta$ by about 30$\%$ to 0.8 at the low frequency end. The
experimental value found by Berg et al is about 1.2.
\par
If a shear rate S is now switched on which result in a mean flow
in (say) the x-direction, then the velocity can be written as \bea
\vec v=Sy\hat{e_{x}} \eea The shear rate S introduces a new length
scale $k_{c}^{-1}$ defined by \bea S=\Gamma k_{c}^{3} \eea Strong
shear implies $k_{c}>\kappa$, while the reverse is the case of
weak shear. Our primary interest will be in strong shear which
will always be the case, sufficiently close to the critical point.
For the shear thinning of a viscoelastic fluid we now have two
dimensionless numbers in the problem
$z_{1}=\frac{\omega}{2\Gamma_{0}\kappa^{3}}$ and
$z_{2}=\frac{\omega}{2\Gamma_{0}k_{c}^{3}}$ and Eq.(6) is
generalized to \bea
\eta=\eta_{0}\kappa^{-x_{\eta}}[G(z_{1},z_{2})]^{-\frac{x_{\eta}}{3}}
\eea The one loop answer for $G(z_{1},z_{2})$ is \bea
G_{1}(z_{1},z_{2})=[1+0.6z_{1}][1+\frac{3}{2z_{2}^{\frac{1}{2}}}\times\frac{1+\frac{0.2}{z_{1}^{\frac{1}{2}}}}{1+\frac{1}{z_{1}}}]^{2}
\eea in the case of strong shear i.e. $k_{c}>\kappa$, which in the
language of $z_{1}$ and $z_{2}$ means $\frac{z_{1}}{z_{2}}>1$.
This is the one loop result.
\par
Using the fact that the one loop scale factor viscoelasticity is
small by a factor of two compared to the experiment, one can use
the improved formula \bea
G(z_{1},z_{2})=[1+1.2z_{1}][1+\frac{3}{2z_{2}^{\frac{1}{2}}}\times\frac{1+\frac{0.2}{z_{1}^{\frac{1}{2}}}}{1+\frac{1}{z_{1}}}]^{2}
\eea The central results of our work are contained in Eqs.(10) and
(12). We now sketch the derivation . The equation of motion for
the order parameter is nonlinear. The effect of the nonlinearity
is to make the transport coefficient divergent in the absence of
the shear. The contribution of the shear to the equation of motion
is a linear term. Consequently, we will work with an effective
equation of motion where the effect of the nonlinear terms will be
handled by a dressing of the transport coefficient. Consequently
in momentum space, the order parameter $\phi(\vec x)$ satisfies
\bea \frac{\partial \phi(\vec k)}{\partial
t}+S\frac{\partial}{\partial k_{y}}\phi(\vec
k)=-\frac{Lk^{2}}{\chi}\phi(\vec k)+\eta \eea where
$\chi^{-1}=k^{2}+\kappa^{2}$ and $L=\Gamma_{0}\chi^{\frac{1}{2}}$
as explained in Eq.(3). We now need to work out the susceptibility
in the presence of S and this is simply retracing Onuki's
calculation with the present L and this leads in a straightforward
fashion to \bea
\chi^{-1}=k^{2}+\kappa^{2}+k_{c}^{2}(\frac{|k_{x}|}{k_{c}}\frac{2}{\pi})^{\frac{1}{2}}
\eea The effect of the changed diffusion coefficient shows up in
the slightly different $k_{x}$ dependence in the last term on the
right hand side. In the expression for viscosity, the
susceptibility will be averaged over all directions and for that
specific case, we will use an angle averaged form of $\chi^{-1}$,
where we replace $|k_{x}|^{\frac{1}{2}}$ by
$k^{\frac{1}{2}}<sin^{\frac{1}{2}}\theta
cos^{\frac{1}{2}}\theta>=\frac{2}{3}k^{\frac{1}{2}}$. The one loop
shear viscosity is now given by \ber
\eta(\kappa,\omega,S)&=&\frac{\eta_{0}}{4\pi}\int d^{3}p
\frac{\chi^{2}(p,\kappa,S)p^{4}}{-i\omega+2\Gamma_{0}p^{2}[\chi(p,\kappa,S)]^{\frac{1}{2}}}\nonumber\\
&=& \frac{\eta_{0}}{4\pi}\int d^{3}p
\frac{p^{4}}{[p^{2}+\frac{2}{3}(\frac{2}{\pi})^{\frac{1}{2}}k_{c}^{2}(\frac{p}{k_{c}})^{\frac{1}{2}}+\kappa^{2}]^{2}}\nonumber\\
& & \times
\frac{1}{[-i\omega+2\Gamma_{0}p^{2}[p^{2}+\frac{2}{3}(\frac{2}{\pi})^{\frac{1}{2}}k_{c}^{2}(\frac{p}{k_{c}})^{\frac{1}{2}}+\kappa^{2}]^{\frac{1}{2}}]}\nonumber\\
&=& \frac{\eta_{0}}{8\pi\Gamma_{0}}\int d^{3}p
\frac{p^{4}}{[p^{2}+\frac{2}{3}(\frac{2}{\pi})^{\frac{1}{2}}(\frac{z_{1}}{z_{2}})^{\frac{1}{2}}p^{\frac{1}{2}}+1]^{2}}\nonumber\\
& & \times
\frac{1}{[z_{1}+p^{2}[p^{2}+\frac{2}{3}(\frac{2}{\pi})^{\frac{1}{2}}(\frac{z_{1}}{z_{2}})^{\frac{1}{2}}p^{\frac{1}{2}}+1]^{\frac{1}{2}}]}\nonumber\\
& & {} \eer It is the characterization of this one loop integral
which has to be carried out. The limit $z_{2}\rightarrow\infty$
(or $k_{c}\rightarrow 0$, i.e. no shear) has already been treated
and this is what we represent to the lowest order by Eq.(7). It
should be noted that Eq.(15) has a logarithmic divergence rather
than a power law divergence. To extract $G(z_{1},z_{2})$ from
Eq.(15), we use the fact that $x_{\eta}\ll 1$ and expand Eq.(10)
as \ber
\eta&=&\eta_{0}\kappa^{-x_{\eta}}[G(z_{1},z_{2})]^{-\frac{x_{\eta}}{3}}\nonumber\\
&\simeq&
\eta_{0}[1-x_{\eta}ln\kappa-\frac{x_{\eta}}{3}ln G(z_{1},z_{2})]\nonumber\\
&=& \eta_{B}+\eta_{0}[ln\frac{\Lambda}{\kappa}-\frac{1}{3}ln
G(z_{1},z_{2})]\nonumber\\
& & {} \eer where $\eta_{B}$ is a background viscosity and the
part within the square brackets emerges from the loop calculation.
\par
For $z_{2}\rightarrow\infty$, Eq.(15) becomes \ber
\eta(\omega,\kappa)&=&
\frac{\eta_{0}}{2\Gamma_{0}}\int_{0}^{\frac{\Lambda}{\kappa}}dp
\frac{p^{6}}{(1+p^{2})^{2}}\frac{1}{[z_{1}+p^{2}(1+p^{2})^{\frac{1}{2}}]}\nonumber\\
&=&
\frac{\eta_{0}}{2\Gamma_{0}}[ln\frac{\Lambda}{\kappa}+ln2-2+\frac{\pi}{4}-\frac{\pi}{16}z_{1}+....]\nonumber\\
& & {} \eer for $z_{1}\ll 1$. On the other hand for $z_{1}\gg 1$,
$\eta(\omega,\kappa)=\frac{\eta_{0}}{2\Gamma_{0}}[ln\frac{\Lambda}{\kappa}-\frac{1}{3}lnz_{1}+.....]$,
which when combined with Eq.(17), leads to the approximation shown
in Eq.(7).
\par
Another limit which can be similarly explored is
$\omega\rightarrow 0$, i.e. $z_{1}\rightarrow 0$ and
$z_{2}\rightarrow 0$, but $\frac{z_{1}}{z_{2}}\neq 0$ in which
case, Eq.(15) becomes \bea
\eta(S,\kappa)=\frac{\eta_{0}}{2\Gamma_{0}}\int_{0}^{\frac{\Lambda}{\kappa}}
dp
\frac{p^{4}}{[p^{2}+\frac{1}{2}(\frac{z_{1}}{z_{2}})^{\frac{1}{2}}p^{\frac{1}{2}}+1]^{\frac{5}{2}}}\eea
For $\frac{z_{1}}{z_{2}}\ll 1$, an expansion similar to Eq.(17)
obtain. For $\frac{z_{1}}{z_{2}}\gg 1$, we scale momenta by
$(\frac{z_{1}}{z_{2}})^{\frac{1}{3}}$ to find \ber
\eta(S,\kappa\rightarrow
0)&=&\frac{\eta_{0}}{2\Gamma_{0}}\int_{0}^{\frac{\Lambda}{k_{c}}}
dp
\frac{p^{4}}{[p^{2}+\frac{1}{2}p^{\frac{1}{2}}+(\frac{z_{2}}{z_{1}})^{\frac{2}{3}}]^{\frac{5}{2}}}\nonumber\\
&\simeq&
\frac{\eta_{0}}{2\Gamma_{0}}[ln\frac{\Lambda}{k_c}+\frac{4}{3}ln2-\frac{8}{3}+\frac{2\pi}{3}+...]\nonumber\\
& & {} \eer Combining Eq.(19) with the small $\frac{z_{1}}{z_{2}}$
form, \bea G(z_{1}\rightarrow 0,z_{2}\rightarrow
0)=[1+0.3(\frac{z_{1}}{z_{2}})^{\frac{1}{2}}]^{2} \eea Finally as
$\kappa\rightarrow 0$, $z_{1}\rightarrow \infty$ with finite
$z_{2}$ and Eq.(15) reduces to \bea
\eta(\omega,S)=\frac{\eta_{0}}{2\Gamma_{0}}\int_{0}^{\frac{\Lambda}{k_{c}}}
dp
\frac{p^{6}}{[p^{2}+\frac{1}{2}p^{\frac{1}{2}}]^{2}[z_{2}+p^{2}(p^{2}+\frac{1}{2}p^{\frac{1}{2}})^{\frac{1}{2}}]}\eea
For $z_{2}\rightarrow 0$, we recover Eq.(19), while for
$z_{2}\rightarrow\infty$ \bea
\eta(\omega,S)=\frac{\eta_{0}}{2\Gamma_{0}}[ln\frac{\Lambda}{k_{c}}-\frac{1}{3}lnz_{2}-\frac{3\pi}{8z_{2}^{\frac{1}{2}}}+...]
\eea We thus obtain the following limiting forms: \ber
G(z_{1},z_{2})&\rightarrow& 1+0.6z_{1}\hspace{1.5cm} if
\hspace{.2cm} z_{2}\rightarrow\infty\nonumber\\
G(z_{1},z_{2})&\rightarrow&
0.6z_{1}[1+\frac{3}{2}\frac{1}{z_{2}^{\frac{1}{2}}}]^{2}
\hspace{0.5cm}if \hspace{0.2cm}
z_{1}\rightarrow\infty\nonumber\\
G(z_{1},z_{2})&\rightarrow&
[1+0.3(\frac{z_{1}}{z_{2}})^{\frac{1}{2}}]^{2} \hspace{0.6cm} if
\hspace{0.2cm} z_{1},z_{2}\rightarrow 0\nonumber\\
& & {} \eer The form of $G(z_{1},z_{2})$ taking into account all
the above constraints is given by Eq.(11) and with the
phenomenological improvement, we arrive at Eq.(12). Very recently
Berg [15] has arrived at a phenomenological rule for shear
thinning. The principle difference between Eq.(11) and Bergs'
observations lie in the existence of the nonanalytic behavior in
$G(z_{1},z_{2})$. This could be an interesting issue during
comparison with experimental data.


\end{document}